**Fail-closed conformal multiresolution error control for conservative three-dimensional dose remapping in synthetic phantoms**

Yuntao Wang, MS

Independent Researcher, Phoenix, Arizona, USA

ORCID iD: 0009-0005-5499-6258 | Correspondence: yuntaowang00@gmail.com

## Abstract

Background: Conservative deformable dose accumulation can be formulated by transporting mass and deposited energy across nonmatching grids. A fine fixed-sampling calculation may reduce reference-relative allocation error but uses a fixed sample count per active source cell regardless of case difficulty, whereas an adaptive calculation requires a principled rule for deciding when a candidate field is sufficiently accurate.

Purpose: To develop and evaluate, under a pre-execution-locked computational protocol, a fail-closed multilevel quasi-Monte Carlo (QMC) release controller for reference-relative numerical error in conservative three-dimensional dose remapping.

Methods: Conformal multiresolution error control (CoMERC) wrapped a conservative replicated nested randomized quasi-Monte Carlo (NQMC) estimator for a fixed cell-centered piecewise-constant source representation and known deformation. Two separately randomized nested replicates were evaluated at a probe level of 4 and candidate levels of 8 and 16. For each candidate level, four endpoints were evaluated: three reference-mass-weighted root-mean-square errors (RMSEs)—global, high-gradient, and density-gradient-threshold—and eligible-cell maximum absolute error. The deployed signal was the endpoint-wise maximum of the inter-resolution discrepancy and a directional same-level replicate discrepancy. A scalar split-conformal multiplier was calibrated jointly over both candidate levels and all endpoints from 240 synthetic cases. The locked controller released level 8 only when all calibrated bounds were within tolerance, otherwise extended both replicates to level 16, and returned ABSTAIN when level 16 also failed. The frozen controller was evaluated once on 400 independent cases from the calibration generator and then descriptively on 120 cases from a prespecified shifted generator. No patient data or manual case review was used.

Results: The noninterpolated rank-229 calibration multiplier was 0.709363. In the locked primary set, joint two-level/four-endpoint coverage was 394/400 (98.50%; one-sided 95% lower limit, 97.06%), 388/400 cases (97.00%; lower limit, 95.18%) received an automatic output, and 0/388 released outputs exceeded any tolerance (one-sided 95% upper limit, 0.769%). Final actions were level 8 for 190 cases, level 16 for 198, and ABSTAIN for 12. Applied to the 400 archived primary records, the mean case-normalized, policy-implied sample ratio was 0.5844, with each case normalized to always executing both randomized production replicates through level 16; the prespecified one-sided 95th-percentile bootstrap upper limit was 0.6194. Operational equivalence of the actual sequential executor was evaluated descriptively in 40 prespecified cases. All seven locked primary gates passed. In the shifted-generator set, 113/120 cases were released, with one tolerance-violating release.

Conclusions: In the locked target generator, CoMERC provided case-dependent reported bounds for reference-relative QMC target-cell allocation error under a marginal finite-sample coverage guarantee. The frozen policy implied lower modeled production-sample work by stopping at level 8 when the calibrated release test passed, while ABSTAIN preserved the rule that no terminal field was released when its bound remained unqualified. The shifted-generator result illustrates that the calibration-based guarantee does not transfer automatically beyond the target population. The study supports a numerical-methods claim, not patient-level, registration-level, or clinical safety validation.

## 1. Introduction

Deformable dose accumulation maps quantities between anatomies or image grids and is used in research and adaptive-radiotherapy workflows. The numerical remapping step is distinct from deformable image registration: even for a fixed deformation, discretization and interpolation can alter local dose and extensive quantities. AAPM Task Group 132 and subsequent dose-accumulation literature emphasize that registration, mapping, and validation errors require explicit characterization rather than visual plausibility alone.[1–8]

Conservative remapping transports mass and deposited energy before recovering dose, thereby separating total-quantity conservation from the spatial allocation of those quantities. Conservative transport has a long history in arbitrary Lagrangian-Eulerian and geophysical remapping, and energy-conserving formulations have recently been developed specifically for dose deformation and dose summation.[9–12] Global conservation, however, is not sufficient for spatial accuracy: an algorithm may preserve total mass and energy while assigning them to incorrect target cells.

A fixed high sampling level provides a simple accuracy strategy but uses the same sample count per active source cell regardless of case difficulty. Multilevel estimators can reuse nested samples and stop earlier, but a stopping rule based only on the difference between successive levels can fail when two biased levels agree. Replicated randomized quasi-Monte Carlo sampling supplies an additional disagreement signal while retaining low-discrepancy integration and nested reuse.[13–15]

Split conformal calibration provides a finite-sample marginal coverage statement under exchangeability without specifying a parametric error model.[16,17] Related distribution-free risk-control methods and recent conformal applications in medical-image registration motivate using calibration not merely to display uncertainty but to control whether an automated numerical result is released.[18–20] The guarantee remains population-specific and marginal; it is not deterministic protection for each individual case and does not automatically extend to shifted generators.

We therefore developed conformal multiresolution error control (CoMERC), a fail-closed release controller around a conservative replicated NQMC estimator. The method calibrates one scalar against a joint family comprising two candidate resolutions and four dose-error endpoints. It releases a candidate only if all calibrated bounds meet prespecified tolerances; otherwise it escalates or abstains. This article reports a fully synthetic, pre-execution-locked confirmatory evaluation, a prespecified shifted-generator stress analysis, and descriptive post-lock executor, fixed-strategy, and scaling validations. The contribution is the release-control architecture and its validation, not a claim of being the first energy-conserving dose-remapping method.

## 2. Materials and Methods

### 2.1 Study design and governance

The study identifier was COMERC-LS073. CoMERC method version 0.72.0 was evaluated under protocol version 0.73. Before any formal production-estimator or numerical-reference outcome was computed, the accepted case manifests, method source, random streams, numerical floors, tolerances, reference-quality rules, statistical analysis, failure handling, and seven conjunctive primary gates were hashed and locked. Geometry generation and deterministic acceptance checks used to construct the manifests occurred before the lock and did not use production-estimator or reference-error outcomes. Formal result directories were empty at lock. Previously used engineering and confirmatory cases were excluded from the new calibration and primary cohorts.

Execution followed the fixed order: 240 calibration cases; freezing of the conformal multiplier; 400 primary evaluation cases; freezing of the primary analysis; 120 descriptive shifted-generator cases; and descriptive executor and scaling validations. Missing, malformed, unexpected, or reference-quality-failing primary records would make the study incomplete rather than reduce the denominator. Successful case records were immutable. No case was manually selected, repaired, excluded, or relabeled.

All data were synthetic. The workflow used no patient or public-patient data, clinical images, treatment-planning system, physical measurement, human-subject activity, or manual scientific scoring. Table 1 summarizes the locked design.

Table 1. Locked confirmatory study design.

| Component | Locked specification |
|---|---|
| Method and protocol | CoMERC method 0.72.0; protocol COMERC-LS073 version 0.73 |
| Calibration | 240 cases from the target generator; noninterpolated split-conformal rank 229 at nominal miscoverage 0.05 |
| Primary evaluation | 400 independent cases from the same target generator; one locked analysis |
| Shifted-generator stress analysis | 120 prespecified case specifications and seeds; execution and analysis occurred after the primary verdict was frozen; descriptive only |
| Numerical estimator | Two separately randomized nested QMC replicates; probe level 4; candidate output levels 8 and 16 |
| Joint error family | Two candidate levels multiplied by four endpoints: reference-mass-weighted global, high-gradient, and density-gradient-threshold RMSE, plus eligible-cell maximum absolute error |
| Release policy | Release level 8 if all four bounds pass; otherwise extend to level 16; release level 16 if all bounds pass; otherwise ABSTAIN |
| Primary inference | Seven conjunctive gates for completeness, joint coverage, family coverage, tolerance-violating released output, release rate, sample work, and numerical-integrity conservation |

*Abbreviations: CoMERC, conformal multiresolution error control; NQMC, nested randomized quasi-Monte Carlo; QMC, quasi-Monte Carlo; RMSE, root-mean-square error.*

### 2.2 Conservative cell-total remapping model

The source domain was discretized into rectilinear cells. Density, dose-like field, and two region-of-interest indicators were evaluated at source-cell centers and treated as piecewise constant within each source cell. For source cell $c$ with volume $V_c$, density $\rho_c$, and dose $d_c$, the extensive mass and deposited energy were

$$m_c = \rho_c V_c, e_c = m_c d_c.$$

Each cell's total mass, energy, target-structure mass, and organ-at-risk structure mass were divided equally among its QMC points. Points were transformed by the case deformation and assigned to target cells; points outside the target domain were accumulated explicitly. Target-cell dose was recovered as $d_k = E_k / M_k$ for cells with positive deposited mass. Thus, QMC approximated the target-cell allocation of cellwise extensive quantities rather than repeatedly evaluating a continuous density or dose function at every sample point.

Mass and energy followed identical point trajectories, so source totals equaled the sum of in-domain and explicitly accumulated out-of-domain totals up to floating-point summation. The prespecified conservation gate was therefore a numerical-integrity and bookkeeping invariant. It did not substitute for spatial dose-error endpoints.

**2.3 Replicated nested randomized quasi-Monte Carlo estimator**

For each active source cell, the production estimator used an unscrambled Sobol base sequence with a cell-specific modulo-one additive shift in the unit cube. Two separately randomized replicates were generated from distinct case-level shift seeds. A level L used the first $L^3$ points per active cell in each replicate. Levels 4, 8, and 16 were nested; escalation retained all prior points. Replicate mass and deposited-energy arrays were averaged separately at each level, and dose was then formed as the ratio of the averaged energy to the averaged mass.

For candidate level $L \in \{8, 16\}$ and endpoint $j$, the inter-resolution discrepancy compared the averaged level-$L/2$ and level-$L$ fields, with the higher-level average supplying reference mass weights and masks. The same-level replicate discrepancy was directional: replicate A was the compared field and replicate B acted as a pseudo-reference supplying weights and masks. The deployed signal was

$$D_{i,L,j} = max\left(D_{i,L,j}^{inter}, D_{i,L,j}^{rep}\right).$$

The same frozen direction and mask construction were used in calibration and evaluation. No post-lock symmetrization or alternative discrepancy was introduced.

**2.4 Synthetic target and shifted generators**

Calibration and primary cases were independent draws from one frozen target generator. It sampled source shapes, spacings, anisotropy, target-grid spacing, density and dose-field variants, phase, fixed-form target and organ-at-risk indicators discretized on randomized grids, and one of four deformation families: affine, coupled sine, radial bump, or twist. During manifest construction, each case allowed at most 50 deterministic deformation-parameter draws from its frozen random stream to meet the 9 × 9 × 9 sampled-Jacobian threshold; exhausting all attempts caused case-generation failure. Prepared cases also had to pass automatic mapping-domain, finite-value, containment, inverse-consistency, and denser sampled positive-Jacobian checks. These finite-lattice checks did not establish global nonfolding or diffeomorphism over the continuous domain. Rejected draws were governed by deterministic code and random streams; there was no manual review.

The 120-case stress cohort came from a separate prespecified shifted generator. It used larger source-grid templates, wider spacing and anisotropy ranges, wider target-to-source spacing ratios, lower accepted sampled-Jacobian floors, larger deformation amplitudes, and composite affine-sine, twist-radial, or double-bump transformations. It was therefore a shifted-generator set, not solely a composite-deformation set. Complete locked ranges and observed cohort summaries are given in Supplementary Tables S1 and S3.

**2.5 Separately implemented numerical reference**

Calibration errors and study outcomes were measured against a separately configured, convergence-screened high-sample QMC reference pathway. The reference did not call the production deposition routine. It used explicit target indexing, a separate deposition path, two separately seeded SciPy Sobol sequences with left linear-matrix scrambling followed by a digital shift, and an additional per-cell modulo-one additive shift generated from a deterministically derived pseudorandom stream. The default fine reference used 65,536 points per active source cell and a 32,768-point coarse comparison. The final fine reference was formed by averaging replicate A and B mass arrays and deposited-energy arrays separately and then computing dose as the ratio of the averaged energy to the averaged mass.

Reference quality used prespecified internal-consistency and convergence screens across all four endpoints. Replicate agreement compared fine replicate A with the combined fine A/B reference formed by separately averaging their mass and deposited-energy arrays and then computing dose;

that combined reference supplied weights and masks. Coarse-to-fine agreement compared coarse replicate A with fine replicate A. Prespecified replicate thresholds were 0.008, 0.016, 0.008, and 0.200 Gy for global RMSE, high-gradient RMSE, density-gradient-threshold RMSE, and maximum error, respectively. Corresponding coarse-to-fine thresholds were 0.015, 0.030, 0.015, and 0.300 Gy. A failing case was automatically recomputed once at 131,072 fine and 65,536 coarse points per active cell. Persistent failure would make that study phase incomplete. These screens were not strict upper bounds on remaining reference error, and residual reference uncertainty was not propagated into the CoMERC bounds. The numerical reference shared the source fields, deformation implementation, grid definitions, and low-level numerical primitives with the production method; it was not an external physical gold standard.

**2.6 Error endpoints and masks**

Three endpoints were reference-mass-weighted RMSEs. For endpoint mask $M_j$ and reference target-cell mass $M_k^{ref}$,

$$RMSE_{w,j} = \left( \frac{\sum_{k \in M_j} M_k^{ref} \left( d_k - d_k^{ref} \right)^2}{\sum_{k \in M_j} M_k^{ref}} \right)^{1/2} .$$

Eligible cells satisfied $M_k^{ref} > 0.005 \, max_r \, M_r^{ref}$. Global RMSE used all eligible cells. Eligible-cell maximum absolute error was $max_{k \in M_{eligible}} |d_k - d_k^{ref}|$.

The high-gradient mask required eligible reference cells with dose at least 10% of the 60-Gy prescription and dose-gradient magnitude at least $0.006 \, D_{Rx} / mm$. If fewer than four cells qualified, the mask used the upper gradient quartile among eligible cells with dose at least 5% of prescription. The density-gradient mask required eligible cells with reference-density-gradient magnitude at least $0.045 \, g \, cm^{-3} mm^{-1}$; if fewer than four cells qualified, the upper quartile among eligible density gradients was used. For realized candidate-versus-reference errors, the high-sample numerical reference supplied weights and masks. For inter-resolution and replicate discrepancies, the higher-level average or replicate B supplied them, respectively. Supplementary Table S2 gives endpoint floors, tolerances, and reference-quality thresholds.

**2.7 Joint conformal calibration and release rule**

For calibration case $i$, candidate level $L$, and endpoint $j$, the normalized ratio was

$$R_{i,L,j} = \frac{E_{i,L,j}}{max\left(D_{i,L,j}, \epsilon_j\right)},$$

where $E_{i,L,j}$ was the realized error relative to the numerical reference and $\epsilon_j$ was the endpoint-specific floor. The case score was

$$S_i = \max_{L \in \{8,16\}, j \in \{1,2,3,4\}} R_{i,L,j}.$$

With $n = 240$ and nominal miscoverage $\alpha = 0.05$, the locked noninterpolated order-statistic rank was $\lceil (n+1)(1-\alpha) \rceil = 229$. The 229th sorted score was frozen as scalar multiplier $q$ before evaluation. For a new case, the reported bound was $B_{L,j} = q\, max\left(D_{L,j}, \epsilon_j\right)$. Level $L$ was released only when $B_{L,j} \le \tau_j$ for all four endpoints. Floors were 0.002, 0.004, 0.002, and 0.020 Gy; tolerances were 0.250, 0.500, 0.250, and 1.500 Gy in the endpoint order above. The controller first tested level 8, then extended both nested replicates to level 16 if needed, and otherwise returned ABSTAIN with no automatic numerical output (Figure 1).

In this article, "fail-closed" refers only to the specified numerical qualification logic: no candidate field was automatically released unless all four prespecified bounds were successfully computed and satisfied their tolerances. It does not denote clinical safety certification or protection against registration, modeling, software-system, or distribution-shift failures.

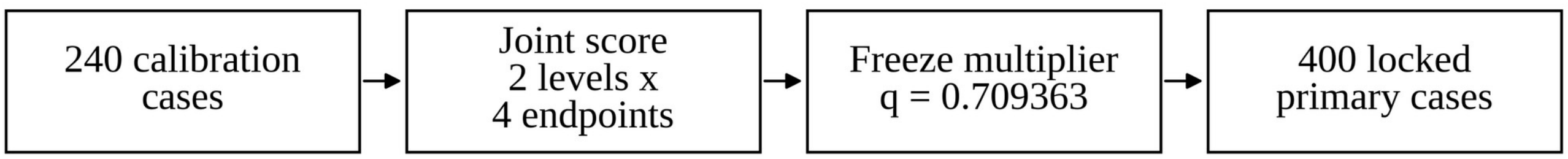

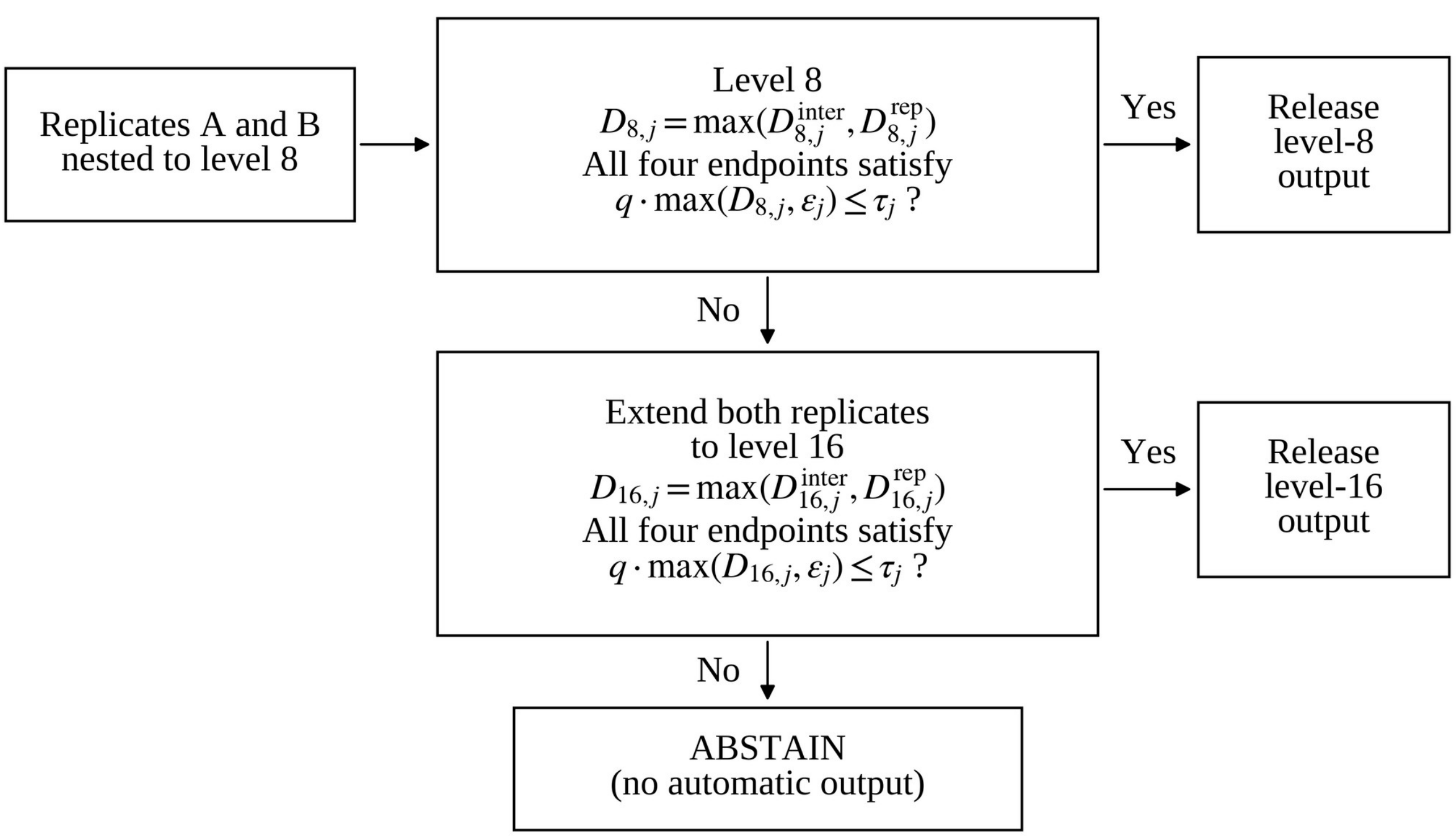

Figure 1. Study separation and fail-closed release policy. (A) The calibration cohort produced one joint multiplier before the locked primary cohort was evaluated. (B) Two separately randomized nested-QMC replicates generated inter-resolution and directional same-level replicate discrepancies. A candidate was released only when all four bounds q·max(DL,j, εj) were within their tolerances; otherwise the controller escalated or abstained.

Split-conformal calibration supports a marginal finite-sample statement for a new exchangeable case drawn from the target generator. Because release requires the selected-level bounds to remain below the numerical tolerances, the construction also controls the marginal event of release together with a tolerance violation. It does not by itself guarantee a 5% or lower tolerance-violation probability conditional on release, deterministic casewise safety, simultaneous protection for an indefinite deployment sequence, or automatic transfer to the shifted generator.

**2.8 Primary estimands and statistical gates**

The primary analysis was the conjunction of seven locked gates: (1) exactly 400 valid evaluation records; (2) joint coverage across both levels and all endpoints with a one-sided 95% Clopper-Pearson lower limit at least 0.90; (3) the same lower limit at least 0.80 within each of four deformation families; (4) among released cases, a one-sided 95% upper limit for any tolerance-violating selected output no greater than 0.025; (5) release-rate lower limit at least 0.85; (6) mean relative sample work no greater than 0.75 and a case-level nonparametric percentile-bootstrap 95th-percentile upper limit strictly below 0.85; and (7) maximum relative mass or energy conservation discrepancy no greater than $10^{-12}$. Exact binomial limits followed Clopper and Pearson.[21] Relative sample work was a policy-implied, case-normalized production-sample ratio computed from the frozen action assigned to each archived full nested record. A level-8 release contributed 0.125 relative work, whereas any case extended through level 16 contributed 1.0, normalized to always executing both randomized production replicates through level 16 for that case. The reported mean was the arithmetic mean of the case ratios, not the ratio of cohort-total sample counts. This quantity did not represent wall-clock work incurred by the confirmatory full-record computations. The bootstrap used 20,000 nonparametric case resamples with frozen PCG64 seed 730200.[22] Wall-clock runtime was descriptive and was not a primary gate.

**2.9 Post-lock descriptive validation**

After the primary result was frozen, the actual sequential executor was rerun on 40 prespecified evaluation cases and compared with stored full nested records. A same-case strategy analysis compared always level 8, CoMERC, and always level 16 using the same 400 cases and estimator family (Supplementary Figure S3 and Table S5). Five prespecified cases formed a fixed-level transport-kernel benchmark with cubic source shapes 8, 12, 16, 20, and 24 and immutable seeds 730301-730305; this original benchmark did not execute the full release controller (Supplementary Table S6). A later post-lock descriptive extension completed level 16 and exercised the release-rule executor for the two largest geometries (Supplementary Table S7). Those extension cases were not evaluated against the high-resolution numerical reference, so no accuracy or transferred conformal-coverage claim was made. These analyses were descriptive and could not change the confirmatory verdict.

### 2.10 Software, automation, and reporting

Case generation, execution, quality control, and analysis were automated. Formal records were atomically written JSON files keyed by immutable case ID. A complete versioned reproducibility archive contains the frozen source, manifests, random seeds, raw case records, official analyses, a separately implemented reanalysis, regression tests, and SHA-256 verification. The current executable paths and validation checks are listed in Supplementary Table S8. For this PDF-only preprint, the archive is available from the author upon request and is intended for deposition in a permanent public repository without changing the frozen scientific files.

OpenAI ChatGPT was used to assist with protocol documentation, code drafting and debugging, automated quality-assurance workflow development, figure preparation, and manuscript drafting and editing. The author retains final responsibility for the study design, source code, analyses, numerical results, references, and posted content.

## 3. Results

### 3.1 Calibration and completion

All 240 calibration cases completed successfully. The 229th noninterpolated order statistic yielded $q = 0.7093632618441885$ (Supplementary Figure S1). Three calibration cases required the single automatic reference-precision escalation. All 400 primary cases and all 120 shifted-generator cases completed; no case had a permanent reference-quality failure. Two shifted-generator cases required reference escalation (Supplementary Table S3).

**3.2 Locked primary analysis**

The complete primary cohort contained 99 affine, 91 coupled-sine, 109 radial-bump, and 101 twist cases. Joint two-level/four-endpoint coverage was 394/400 (98.50%), with one-sided 95% lower limit 97.06%. A numerical output was released for 388/400 cases (97.00%), with lower limit 95.18%. Among released outputs, 0/388 exceeded any selected-level tolerance; the one-sided 95% upper limit was 0.769%.

Level 8 was released for 190 cases, level 16 for 198, and 12 cases returned ABSTAIN. Applied to the 400 archived primary records, the arithmetic mean of the case-normalized, policy-implied sample ratios was 0.584375. The prespecified 20,000-resample one-sided 95th-percentile bootstrap upper limit was 0.619375. The maximum relative mass or energy conservation discrepancy across method and reference fields was $6.62 \times 10^{-15}$. Every locked gate passed (Table 2 and Figure 2).

Table 2. Conjunctive locked primary gates.

| Gate | Locked criterion | Observed result | Verdict |
|---|---|---|---|
| Completeness | Exactly 400 valid primary records; no permanent reference-quality failure | 400/400 valid; 0 permanent failures | PASS |
| Joint coverage | One-sided 95% lower limit >= 90% | 394/400 (98.50%); lower 97.06% | PASS |
| Family empirical coverage | One-sided 95% lower limit >= 80% in every family | Family lower limits 93.78%-95.39% | PASS |
| Tolerance-violating released output | One-sided 95% upper limit <= 2.5% | 0/388; upper 0.769% | PASS |
| Release rate | One-sided 95% lower limit >= 85% | 388/400 (97.00%); lower 95.18% | PASS |
| Sampling efficiency | Mean <= 75%; bootstrap upper < 85% | Mean 58.44%; upper 61.94% | PASS |
| Numerical-integrity conservation | Maximum relative discrepancy <= 1e-12 | 6.62 x 10^-15 | PASS |

**A Coverage and release rate**

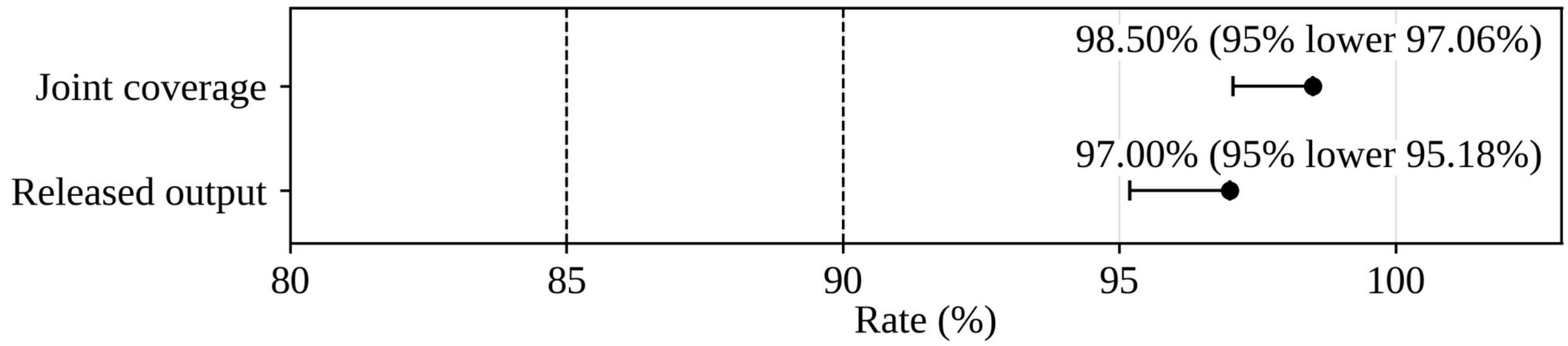

Dashed lines: coverage gate 90%; release gate 85%.

**B Tolerance-violation control**

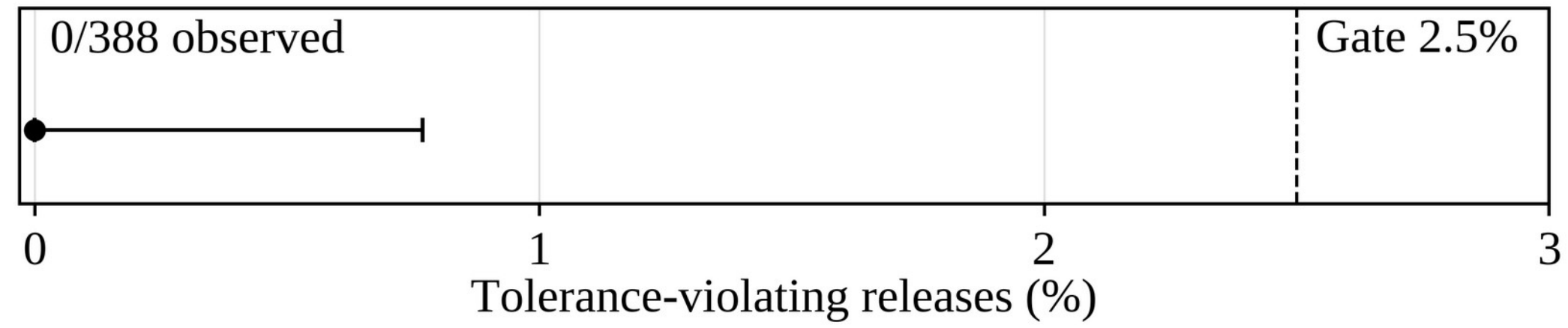

One-sided 95% upper limit: 0.77%.

**C Final actions**

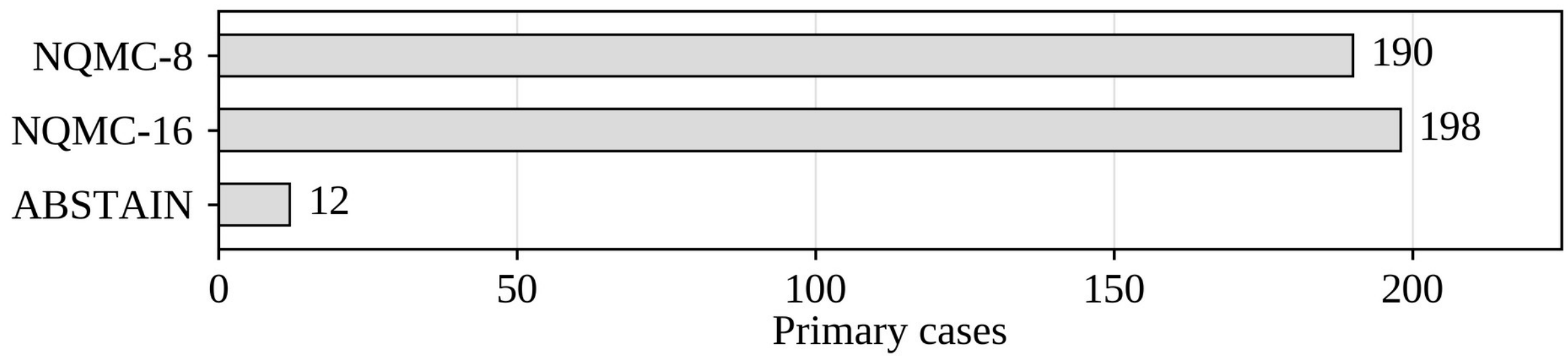

**D Algorithmic sample work**

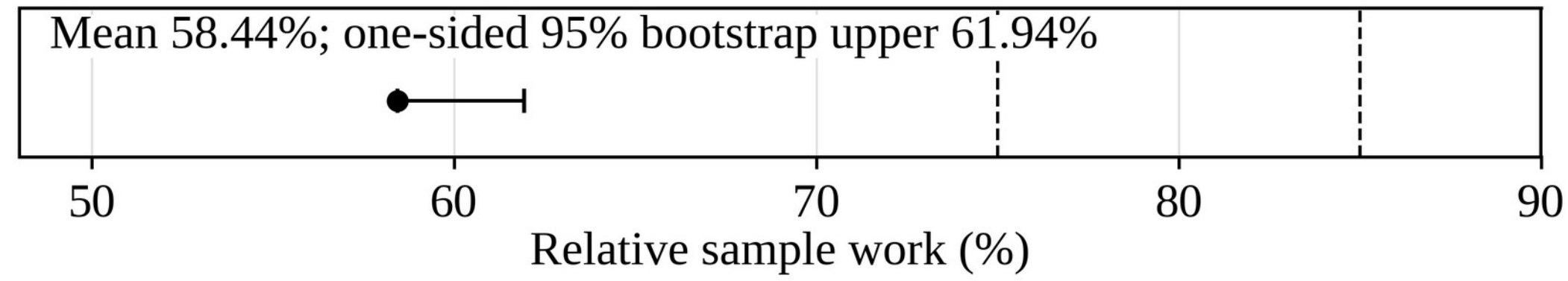

Figure 2. Locked primary results. (A) Joint coverage and release rate with one-sided 95% lower limits; vertical dashed lines mark the 90% joint-coverage and 85% release-rate gates. (B) Tolerance-violating released outputs with the one-sided 95% upper limit and locked 2.5% gate. (C) Final actions. (D) Policy-implied mean relative sample work and the prespecified one-sided 95th-percentile bootstrap upper limit; dashed lines show the mean and bootstrap gates. NQMC, nested randomized quasi-Monte Carlo.

### 3.3 Family empirical coverage and shifted-generator stress analysis

All four target-generator deformation families exceeded the prespecified empirical family-coverage gate. Joint coverage was 97/99 for affine, 90/91 for coupled sine, 107/109 for radial bump, and 100/101 for twist. Their one-sided lower limits ranged from 93.78% to 95.39%. Release rates ranged from 95.05% to 98.17% (Figure 3A and Supplementary Table S4). These subgroup results were empirical diagnostics; the pooled conformal multiplier did not create a separate family-conditional coverage guarantee.

The shifted-generator set was analyzed only after the primary result was frozen. Joint coverage was 116/120 (96.67%), 113/120 cases (94.17%) were released, and seven abstained. One of the 113 released outputs exceeded a tolerance (0.885%); mean relative work was 0.628125 (Figure 3B-C and Supplementary Table S4). Because these cases came from a prespecified shifted generator, the target-population conformal guarantee did not apply. The single event illustrates the absence of deterministic casewise transfer protection but does not by itself establish degradation caused by distribution shift.

**A By deformation family**

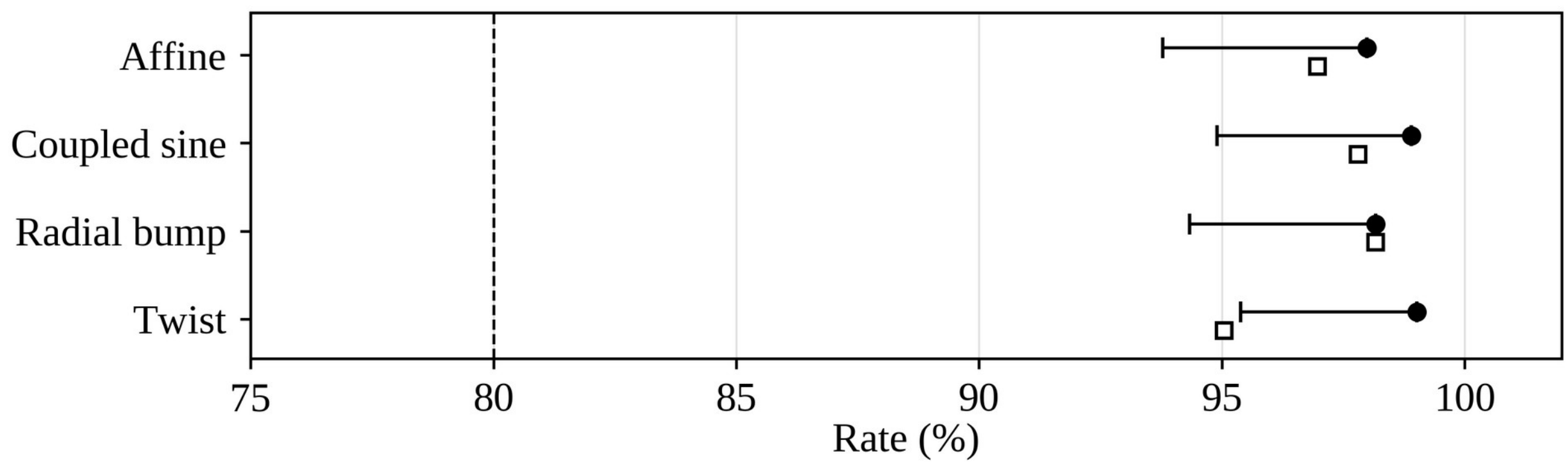

Filled circles: joint coverage with one-sided 95% lower limits.
Open squares: released-output rate; dashed line: 80% gate.

**B Primary vs shifted**

Rate (%)
100
90
80
98.5
96.7
97.0
94.2
Joint coverage
Released output

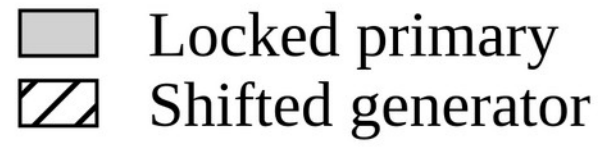

**C Violations among released**

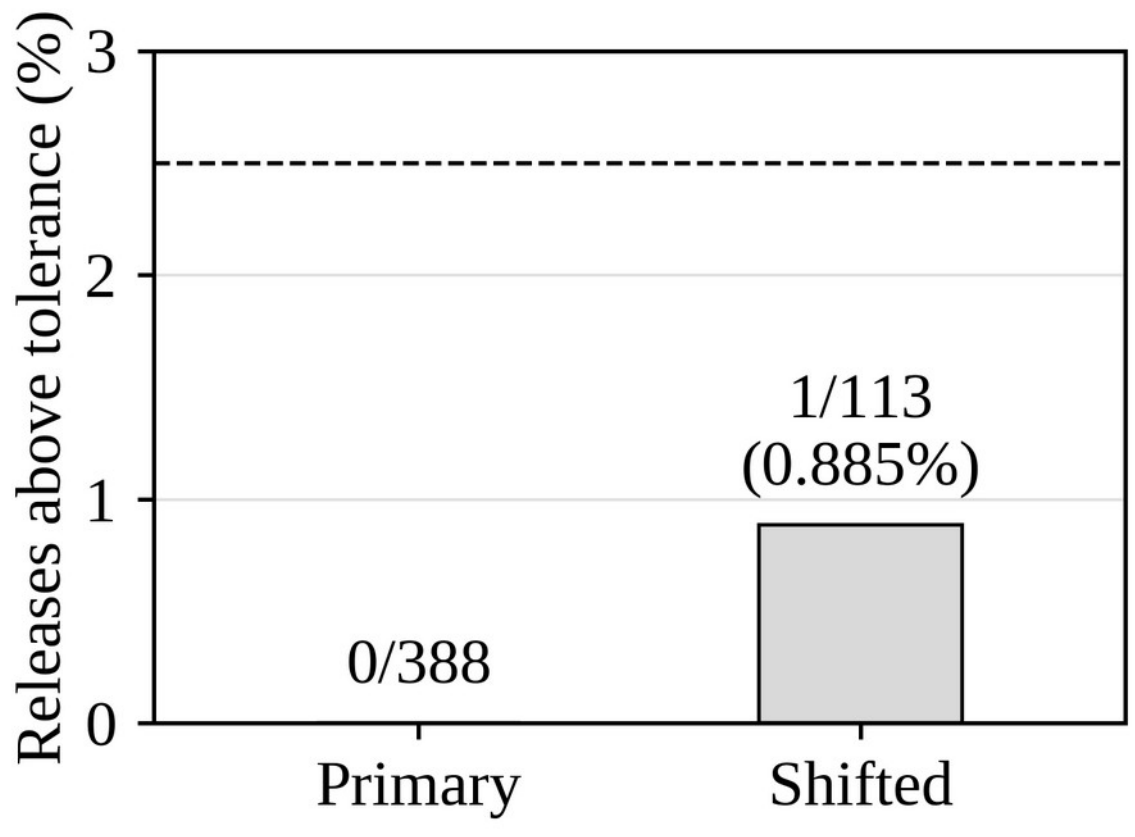

Dashed line: primary gate only (2.5%).

Figure 3. Deformation-family and shifted-generator results. (A) Target-generator family-specific joint coverage with one-sided 95% lower limits and released-output rates. (B) Aggregate joint coverage and release rate in the locked primary and shifted-generator cohorts. (C) Tolerance-violating released outputs, displayed on a 0%-3% scale; the dashed line is the locked primary tolerance-violation gate, which was not an inferential gate for the shifted cohort.

### 3.4 Selected-output errors and post-lock validations

Across the 388 released primary outputs, observed endpoint errors remained within their respective tolerances (Supplementary Figure S2). The 40-case actual-executor validation reproduced all stored actions, arrays, discrepancy signals, and bounds exactly, with maximum numerical difference zero. A separately implemented statistical reanalysis reproduced the conformal rank, multiplier, case decisions, confidence limits, bootstrap result, and official case table.

Prespecified secondary mass-weighted synthetic structure-dose summaries were reconstructed from the archived raw records without rerunning or retuning the study. Among the 388 released primary outputs, median (95th percentile; maximum) absolute error was 0.0208 (0.1497; 0.4847) Gy for mass-weighted synthetic target D95-like and 0.0314 (0.2051; 0.5105) Gy for mass-weighted synthetic organ-at-risk D2-like. Target and organ-at-risk mean-dose summaries, fixed-level comparators, and shifted-generator results are reported in Supplementary Table S9. These descriptive digital-structure metrics are not standard volume-based clinical DVH endpoints, clinical action thresholds, or patient-level validation.

The same-case strategy analysis showed that always level 8 had 22/400 tolerance-violating outputs at relative work 0.125, CoMERC released 388 in-tolerance outputs and abstained on 12 at relative work 0.584375, and always level 16 had 400/400 in-tolerance outputs at relative work 1.0. Among 210 cases escalated beyond level 8, 188 had a level-8 field that was retrospectively within tolerance; all 12 abstained cases had a retrospectively in-tolerance level-16 field. These post-lock findings quantify the controller’s conservatism and were not used to tune it.

The prespecified fixed-level transport-kernel benchmark completed its frozen configured maximum level: level 16 for the 8 × 8 × 8, 12 × 12 × 12, and 16 × 16 × 16 source shapes, and level 8 for the 20 × 20 × 20 and 24 × 24 × 24 cases. The original benchmark did not execute the complete release controller, and the latter two runs were capped at level 8 by the machine-readable manifest; transport-call wall times are therefore not like-for-like level-16 comparisons. A separate post-lock descriptive extension subsequently ran level 16 and the release-rule executor for those two geometries and seeds. Both cases failed the level-8 release rule only on the eligible-cell maximum-error bound and passed all four level-16 release-rule bounds. Because no high-resolution numerical reference was computed for the extension, these outcomes support implementation feasibility only, not numerical accuracy or transferred conformal validity. Executor wall times were 6.97 and 11.77 s, with process-lifetime maximum resident memory of 146.97 and 152.45 MiB, respectively (Supplementary Tables S6 and S7).

### 3.5 Post-lock controller-mechanism diagnostics

These diagnostics were computed only after the confirmatory verdict had been frozen and were not used to alter the score, thresholds, or policy. Under the archived driver-assignment rule, the eligible-cell maximum absolute-error endpoint was the primary score driver in 184/400 primary cases, participated in 190/210 level-8 escalations, and was the sole level-16 trigger in all 12 abstentions. Inter-resolution discrepancy exceeded the directional same-level replicate discrepancy in 3,161/3,200 primary level-endpoint comparisons; the endpoint-specific denominator floors were never active across 6,080 formal calibration, primary, and shifted level-endpoint records.
The density-gradient mask covered a median 95.89% of eligible primary cells, and density-gradient-threshold RMSE was strongly correlated with global RMSE (Pearson r = 0.97065 at level 8 and 0.97267 at level 16), indicating substantial endpoint redundancy in this generator. Six primary cases missed joint conformal coverage, but none produced a tolerance-violating selected output. All 12 abstained level-16 fields were retrospectively within tolerance. These findings are descriptive and are detailed in Supplementary Tables S10-S12.

## 4. Discussion

The locked target-generator evaluation met all seven conjunctive primary gates. The main result is not merely a low average error: the controller calibrated a single joint multiplier across two candidate levels and four endpoints, released 97% of cases, and observed no tolerance-violating release in 388 outputs. The mean of the case-normalized, policy-implied production-sample ratios was 0.5844, with each case normalized to always executing both randomized production replicates through level 16. The terminal ABSTAIN rule prevented the architecture from treating the highest available level as automatically qualified.

The method addresses a specific failure mode identified in an earlier, separately locked study. That predecessor certified the lower level but accepted an uncertified terminal level after escalation; two terminal-level failures remained even after retrospective lower-level threshold tightening. The present controller instead applies the same calibrated release test at levels 8 and 16 and returns no automatic output when the terminal test fails. The earlier negative result remains development evidence and is not pooled with the current inference.

CoMERC complements rather than replaces conservative dose-remapping methods. Energy-conserving dose deformation and summation have been proposed previously, and the present estimator likewise transports extensive mass and energy.[11,12] The new element is a statistical release layer that combines inter-resolution disagreement with replicate instability, calibrates their relation to high-resolution numerical-reference error, and permits explicit abstention. The conservation gate primarily verifies bookkeeping integrity; local RMSE and maximum-error endpoints assess spatial allocation.

The split-conformal statement should be interpreted narrowly. Under exchangeability between calibration cases and a future case from the target generator, the joint score construction yields marginal finite-sample coverage for the complete two-level/four-endpoint family. The reported bound is case-dependent because it uses that case's discrepancies, but the guarantee is not conditional on each released case. Release plus a tolerance violation is controlled as a marginal event; the construction does not by itself prove that the tolerance-violation probability conditional on release is at most 5%. The 0/388 count is empirical evidence from the locked cohort with an exact upper confidence limit, not proof of zero conditional risk.

The shifted-generator analysis reinforces this boundary without supporting an exaggerated causal claim. The shifted cohort changed grid distributions, geometric ranges, Jacobian acceptance, deformation amplitudes, and transformation composition. Its single tolerance-violating release shows that release can fail outside the calibration population. Because the shifted cohort had no prespecified inferential degradation test and the observed 0.885% tolerance-violation fraction was small, the event does not alone demonstrate statistically significant worsening caused by distribution shift.

The same-case strategy comparison clarifies the accuracy-work-abstention trade-off. Always level 8 used one eighth of the modeled sample work of always level 16 but produced 22 tolerance-violating fields. Escalation intercepted all 22 of those observed level-8 violations. CoMERC occupied the middle in work and released 388 in-tolerance outputs. Its 12 terminal abstentions preserved the fail-closed qualification rule, but every corresponding level-16 field was retrospectively within tolerance; terminal abstention therefore did not prevent an observed level-16 violation in this cohort. The 188 retrospectively unnecessary escalations further show conservative over-escalation. Reducing this conservatism would require a newly specified score or policy followed by fresh calibration and evaluation; the locked primary cases cannot be reused for tuning.

This study has several limitations. First, it used compact synthetic phantoms rather than patient anatomy. The primary source grids had 82-168 active source cells and 1,320-4,446 target cells; even the largest post-lock extension had 5,640 active source cells and 29,568 target cells. The benchmarks therefore describe implementation feasibility, not clinical throughput. Second, the controlled quantity was QMC target-cell allocation/deposition error for a fixed cell-centered piecewise-constant source representation, fixed target grid, and known deformation. Source- and target-grid discretization, analytic-field-to-cell-center approximation, image-registration error, clinical dose-calculation error, contour uncertainty, and serial multi-fraction accumulation were not evaluated. Third, the numerical reference used a separate deposition path and randomization configuration but shared the deformation, grids, source fields, metric utilities, and low-level numerical primitives; its internal consistency screens were not strict error bounds, and residual reference uncertainty was not propagated. Fourth, tolerances, floors, and gate thresholds were pre-execution engineering specifications inherited from consumed development work; they defined a falsifiable numerical benchmark but were not clinically validated or claimed to be optimal. Finally, sample work is a modeled production-evaluation count rather than a hardware-independent speedup, and the conformal guarantee does not extend automatically to arbitrary deformation fields, continuous deployment sequences, or shifted clinical populations.

The fully synthetic design also has strengths. It allowed known programmatic fields, automatic sampled positive-Jacobian, finite-value, containment, and inverse-consistency checks, immutable seeds, complete failure accounting, and a large independent calibration/evaluation separation without patient privacy constraints. Synthetic digital phantoms are established tools for controlled imaging research, although they cannot replace clinical validation.[23] Direct comparisons of dose mapping with energy/mass transfer and a broader review of dose mapping and accumulation further contextualize the present fixed-deformation numerical scope.[24,25] The versioned reproducibility archive permits exact statistical reconstruction from all 760 case records and provides a full from-scratch computation entry point.

## 5. Conclusions

A fail-closed conformal release controller governed a replicated nested-QMC conservative remapping estimator in the locked target generator. It provided marginally calibrated, case-dependent bounds for reference-relative QMC target-cell allocation error under a fixed cell-centered source representation and known deformation. The controller released 388/400 primary outputs with no observed tolerance violation. The mean of the case-normalized, policy-implied production-sample ratios was 0.5844, with each case normalized to always executing both randomized production replicates through level 16. ABSTAIN preserved terminal qualification, although all 12 abstained level-16 fields were retrospectively within tolerance. The result supports a reproducible numerical-methods claim within the target generator and should not be interpreted as patient-level, registration-level, or clinical safety validation.

**Data and Code Availability**

A complete versioned reproducibility archive is available from the author upon request (yuntaowang00@gmail.com) and is intended for deposition in a permanent public repository. It contains frozen source code, case manifests, random seeds, all 760 formal case records, official analyses, validation records, post-lock descriptive diagnostics, and package-verification scripts. The archive includes an internal SHA-256 manifest and VERIFY_PACKAGE.py. Any later repository record will preserve the frozen scientific files and will be cited in an updated preprint version.

**Ethics Statement**

No human participants, patient data, public-patient data, animals, clinical images, or identifiable information were used. Institutional review board review and informed consent were not applicable.

**Supplementary Material**

**Fail-closed conformal multiresolution error control for conservative three-dimensional dose remapping in synthetic phantoms**

## S1. Study lineage, lock, and reporting boundary

The current study, COMERC-LS073, evaluated CoMERC method version 0.72.0 under protocol version 0.73. Earlier ACEMR and CoMERC development cohorts, including the separately locked negative COMERC-LS053 study, were consumed before the current calibration and primary manifests were generated. They were not reused in current confidence intervals or thresholds.

At lock, 123 controlled files were hashed and formal calibration, evaluation, and stress result directories were empty. The immutable manifests defined 240 calibration cases, 400 primary cases, 120 shifted-generator stress cases, 40 actual-executor validation cases, and five scaling cases. Geometry generation and deterministic acceptance checks used to construct the accepted case manifests occurred before the lock and did not use production-estimator or numerical-reference error outcomes. Post-lock changes to code, manifests, random seeds, endpoints, floors, tolerances, reference rules, primary gates, and failure rules were prohibited. Execution or environment interruption could be retried only with the same case, code, and random streams; scientific reference-quality failure beyond the single prespecified precision escalation could not be retried.

## S2. Conservative estimator and nested sampling

Source density, dose, and two structure indicators were evaluated at source-cell centers and treated as piecewise constant within each source cell. Cell-total mass, energy, target-structure mass, and organ-at-risk structure mass were distributed equally across QMC points and deposited by transformed target-cell membership. Out-of-domain quantities were accumulated explicitly. Dose was recovered from deposited energy divided by deposited mass.

Production sampling used an unscrambled Sobol base sequence, cell-specific modulo-one additive shifts in the unit cube, two separately randomized replicates generated from distinct case-level shift seeds, and nested levels 4, 8, and 16. A level L used $L^3$ points per active source cell per replicate. When the policy escalated, samples already evaluated at lower levels were retained. At each level, replicate mass and deposited-energy arrays were averaged separately, and dose was formed as the ratio of the averaged energy to the averaged mass.

For candidate level $L$, the inter-resolution discrepancy compared averaged level-$L/2$ and level-$L$ fields, using the level-$L$ average as the pseudo-reference that supplied mass weights and masks. The same-level replicate discrepancy compared replicate A against replicate B, with replicate B supplying weights and masks. The directional definition was frozen before calibration. The deployed endpoint signal was the maximum of the two discrepancies.

**S3. Numerical reference and quality control**

The high-resolution numerical reference used a separate deposition and target-indexing path, NumPy bincount reduction, two separately seeded SciPy Sobol sequences with left linear-matrix scrambling followed by a digital shift, and an additional per-cell modulo-one additive shift generated from a deterministically derived pseudorandom stream. It did not call the production deposition routine. Initial fine and coarse resolutions were 65,536 and 32,768 points per active source cell, respectively. Replicate-agreement screening compared fine replicate A with the combined fine A/B reference formed by separately averaging their mass and deposited-energy arrays and then computing dose; that combined reference supplied weights and masks. Coarse-to-fine screening compared coarse replicate A with fine replicate A. The same combined fine A/B field was used as the final numerical reference. A case failing either screen was automatically recomputed once at 131,072 fine and 65,536 coarse points per active cell. Persistent failure made the applicable study phase incomplete. These internal consistency screens were not strict upper bounds on residual reference error.

The reference used a separately configured deposition path and randomization scheme but shared the synthetic source fields, deformation implementation, source/target grids, metric utilities, QMC integration principle, and low-level numerical primitives with the production method. It should therefore be interpreted as a convergence-screened numerical reference rather than an independent external implementation or physical gold standard.

**S4. Synthetic generators**

The target generator sampled source-grid templates, base spacing, anisotropy, target-spacing ratio, phase, density-field variant, dose-field variant, fixed-form target and organ-at-risk indicators discretized on randomized grids, and one of four deformation families. During manifest construction, each case permitted at most 50 deterministic deformation-parameter draws from its frozen random stream to satisfy the $9 \times 9 \times 9$ sampled-Jacobian threshold (minimum 0.45 for target-generator cases and 0.25 for shifted-generator cases); exhaustion caused generation failure. Prepared cases then underwent automatic finite-value, mapping-domain, containment, sampled positive-Jacobian, and inverse-consistency checks, including an $11 \times 11 \times 11$ Jacobian lattice and a maximum inverse-consistency tolerance of $2 \times 10^{-6}$ mm. These finite-lattice checks did not prove global nonfolding or diffeomorphism over the continuous domain. The shifted generator changed multiple components simultaneously and is more accurately described as a shifted-generator stress population than as only a composite-deformation population. Table S1 reports frozen ranges. All acceptance and rejection decisions were automatic.

Table S1. Frozen synthetic-generator ranges.

| **Parameter** | **Calibration and primary target generator** | **Shifted-generator stress set** |
|---|---|---|
| Source-grid templates | (5,6,7), (6,6,6), (7,6,5), (7,7,6), (8,6,7), (8,8,6) | (7,8,9), (8,8,8), (9,7,8), (9,9,7) |
| Base source spacing | 2.0-3.8 mm | 2.0-4.2 mm |
| Axis anisotropy factor | 0.82-1.18 | 0.65-1.35 |
| Target/source spacing ratio | 0.78-1.18 | 0.65-1.35 |
| Phase parameter | 0.12-0.88 | 0.12-0.88 |

| Parameter | Calibration and primary target generator | Shifted-generator stress set |
|---|---|---|
| Deformation families | affine, coupled sine, radial bump, twist | affine+sine, twist+radial, double radial bump |
| Affine rotation | 0.08-0.32 rad | up to 0.50 rad |
| Affine shear | -0.08 to 0.08 | -0.14 to 0.14 |
| Affine scale | 0.92-1.08 | 0.86-1.15 |
| Twist maximum angle | 0.20-0.55 rad | up to 0.90 rad |
| Radial-bump center | within 18% of domain extent from center | same component rule in composite draws |
| Radial-bump width | 22%-40% of domain scale | same component rule in composite draws |
| Radial displacement | 3.5%-8.5% of minimum extent | up to 13% of minimum extent |
| Coupled-sine amplitude | 0.8%-1.8% of minimum extent | up to 3.0% of minimum extent |
| Translation offset | up to 0.12 source-cell spacing per axis | up to 0.12 source-cell spacing per axis |
| Sampled-Jacobian floor | 0.45 | 0.25 |
| Density and dose variants | eight density variants; eight synthetic dose variants; 60-Gy prescription scale | same field libraries with shifted geometry/grid draws |

*Abbreviations: Gy, gray.*

**S5. Endpoint implementation**

Eligible target cells had reference target-cell mass greater than 0.5% of the maximum reference target-cell mass in that case. Global, high-gradient, and density-gradient-threshold RMSE were weighted by reference target-cell mass. The locked record field for the third endpoint was named interface RMSE, but its implemented mask was defined by a density-gradient threshold. Maximum absolute error was computed among eligible cells only.

The high-gradient mask required reference dose at least 10% of prescription and gradient magnitude at least 0.006 times prescription per millimeter. If fewer than four cells qualified, the upper gradient quartile among eligible cells with dose at least 5% of prescription was used. The density-gradient mask required reference-density-gradient magnitude at least 0.045 g $cm^{-3}$ $mm^{-1}$; if fewer than four cells qualified, the eligible-cell density-gradient upper quartile was used. For discrepancy calculations, the pseudo-reference field for each comparison supplied the weights and masks.

Table S2. Endpoint, conformal-floor, tolerance, and reference-quality definitions.

| Endpoint | Scoring set and weighting | Floor / release tolerance (Gy) | Reference replicate / coarse-fine QC (Gy) |
|---|---|---|---|
| Global RMSE | Reference-mass-weighted over eligible cells | 0.002 / 0.250 | 0.008 / 0.015 |
| High-gradient RMSE | Reference-mass-weighted over deterministic high-gradient mask | 0.004 / 0.500 | 0.016 / 0.030 |
| Density-gradient-threshold RMSE | Reference-mass-weighted over deterministic density-gradient mask | 0.002 / 0.250 | 0.008 / 0.015 |
| Maximum absolute error | Maximum among eligible target cells | 0.020 / 1.500 | 0.200 / 0.300 |

*Abbreviations: QC, quality control; RMSE, root-mean-square error.*

**S6. Conformal and statistical procedures**

The calibration ratio was $R_{i,L,j} = E_{i,L,j} / max\left(D_{i,L,j}, \epsilon_j\right)$, and each case score was the maximum across candidate levels 8 and 16 and all four endpoints. With 240 calibration cases and nominal miscoverage 0.05, the noninterpolated split-conformal rank was 229. The scalar 229th score was frozen before primary execution.

Exact one-sided 95% Clopper-Pearson limits were used for binomial gates. The primary work bootstrap sampled the 400 cases with replacement 20,000 times using PCG64 seed 730200 and reported the 95th percentile as the one-sided upper limit. No post-lock multiplicity adjustment or threshold search was performed. Secondary and post-lock comparisons were descriptive.

**S7. Completion, reference quality, and cohort scale**

Table S3. Dataset completion, numerical-reference quality, and observed grid scale.

| Dataset | Successful / locked | Automatic reference escalation | Permanent reference failure | Observed scale |
|---|---|---|---|---|
| Calibration | 240/240 | 3 | 0 | 82-168 active source cells; 1210-4420 target cells; 88-465 eligible cells |

| Dataset | Successful / locked | Automatic reference escalation | Permanent reference failure | Observed scale |
|---|---|---|---|---|
| Primary | 400/400 | 0 | 0 | 82-168 active source cells; 1320-4446 target cells; 109-497 eligible cells |
| Shifted generator | 120/120 | 2 | 0 | 196-233 active source cells; 1872-6615 target cells; 186-717 eligible cells |

**S8. Family and shifted-generator outcomes**

Table S4. Deformation-family and shifted-generator outcomes.

| Cohort or family | n | Joint coverage | One-sided 95% lower limit | Released output | Tolerance violations among released |
|---|---|---|---|---|---|
| Affine | 99 | 97/99 (97.98%) | 93.78% | 96/99 (96.97%) | 0 |
| Coupled sine | 91 | 90/91 (98.90%) | 94.89% | 89/91 (97.80%) | 0 |
| Radial bump | 109 | 107/109 (98.17%) | 94.34% | 107/109 (98.17%) | 0 |
| Twist | 101 | 100/101 (99.01%) | 95.39% | 96/101 (95.05%) | 0 |
| Shifted-generator aggregate | 120 | 116/120 (96.67%) | Descriptive | 113/120 (94.17%) | 1/113 (0.885%) |

**S9. Same-case strategy comparison**

The post-lock comparison used all 400 locked primary cases and the same stored NQMC estimator outputs. Always NQMC-8 treated the level-8 mean field as the output for every case; always NQMC-16 used the level-16 mean field for every case; CoMERC used the frozen release decisions. Relative work was normalized within each case to always executing both randomized production replicates through level 16; reported means gave equal weight to cases. Because this analysis was conducted after the primary result was known, it is descriptive and cannot support a new confirmatory superiority claim.

Table S5. Post-lock same-case strategy trade-off.

| Strategy | Automatic outputs within tolerance | Automatic outputs above tolerance | ABSTAIN | Mean relative sample work |
|---|---|---|---|---|
| Always NQMC-8 | 378/400 | 22/400 | 0 | 0.125000 |
| CoMERC | 388/400 | 0/388 released | 12 | 0.584375 |
| Always NQMC-16 | 400/400 | 0/400 | 0 | 1.000000 |

*Abbreviations: CoMERC, conformal multiresolution error control; NQMC, nested randomized quasi-Monte Carlo.*

Among the 210 cases escalated beyond level 8 by CoMERC, 188 level-8 arrays were retrospectively within tolerance. All 12 abstained cases had a retrospectively in-tolerance level-16 array. These observations quantify conservative release behavior and were not used to alter the policy.

**S10. Scaling benchmark**

Five prespecified scaling cases were executed with immutable seeds 730301-730305 as a fixed-level transport-kernel benchmark. The original benchmark did not execute the complete release controller. The frozen machine-readable manifest specified maximum level 16 for source shapes 8 × 8 × 8, 12 × 12 × 12, and 16 × 16 × 16, and maximum level 8 for 20 × 20 × 20 and 24 × 24 × 24. The benchmark recorded active source cells, target cells, cumulative sample evaluations through the configured maximum, transport-call wall time, and process-lifetime maximum resident set size. Timings are environment-specific, are not directly comparable across rows with different configured maximum levels, and do not represent end-to-end or clinical throughput.

Table S6. Prespecified descriptive scaling benchmark.

| Source shape / configured max | Active / target cells | Level-8 evaluations | Level-16 evaluations | Total wall time to max (s) | Maximum RSS (MiB) |
|---|---|---|---|---|---|
| 8 × 8 × 8 / L16 | 200 / 2,730 | 204,800 | 1,638,400 | 0.344 | 141.68 |
| 12 × 12 × 12 / L16 | 696 / 6,137 | 712,704 | 5,701,632 | 1.143 | 140.97 |

| **Source shape / configured max** | **Active / target cells** | **Level-8 evaluations** | **Level-16 evaluations** | **Total wall time to max (s)** | **Maximum RSS (MiB)** |
|---|---|---|---|---|---|
| 16 × 16 × 16 / L16 | 1,672 / 11,592 | 1,712,128 | 13,697,024 | 2.629 | 142.24 |
| 20 × 20 × 20 / L8 | 3,232 / 18,900 | 3,309,568 | not attempted | 0.692 | 142.41 |
| 24 × 24 × 24 / L8 | 5,640 / 29,568 | 5,775,360 | not attempted | 1.268 | 144.47 |

*Abbreviations: NQMC, nested randomized quasi-Monte Carlo; RSS, resident set size. Rows end at their originally configured maximum level and are not like-for-like level-16 timing comparisons.*

**S11. Post-lock level-16 scaling extension**

The original prespecified scaling manifest capped the 20 × 20 × 20 and 24 × 24 × 24 cases at level 8. After the primary result had been frozen, a separately labeled descriptive extension used the same geometries and seeds (730304 and 730305) to complete level 16 and exercise the release-rule executor. Both cases failed only the eligible-cell maximum-error release rule at level 8, escalated, and passed all four level-16 release-rule bounds. No high-resolution numerical reference was computed for these extension cases, so the extension supports implementation feasibility only and makes no numerical-accuracy or transferred conformal-coverage claim. It did not modify or replace the original benchmark and was not part of any confirmatory gate.

**Table S7. Post-lock level-16 release-rule executor extension.**

| **Source shape** | **Active / target cells** | **Level-16 evaluations** | **Level-8 release failure** | **Level-16 decision** | Release-rule executor time / maximum RSS |
|---|---|---|---|---|---|
| 20 × 20 × 20 | 3,232 / 18,900 | 26,476,544 | Eligible-cell maximum bound | All four bounds passed; release | 6.973 s / 146.97 MiB |
| 24 × 24 × 24 | 5,640 / 29,568 | 46,202,880 | Eligible-cell maximum bound | All four bounds passed; release | 11.766 s / 152.45 MiB |

*Abbreviations: RSS, resident set size. Timings are environment-specific descriptive implementation measurements, not clinical-throughput estimates.*

**S12. Post-lock executor and archive validation**

**Table S8. Reproducibility and validation checks.**

| Check | Result | Current evidence path |
|---|---|---|
| Actual early-stopping executor | 40/40 actions and arrays exact; maximum difference 0 | CoMERC_LS073_v0.73_Execution/ validation/ ACTUAL_EXECUTOR_ SUBSET_40_SUMMARY.json |
| Separately implemented statistical reanalysis | Rank, q, decisions, exact intervals, bootstrap, and summaries reproduced | CoMERC_LS073_v0.73_Execution/ validation/ INDEPENDENT_REANALYSIS.json |
| Official raw-record reanalysis | Primary analysis JSON and case CSV reproduced byte for byte | CoMERC_LS073_v0.73_Execution/ REPRODUCE_ANALYSES_FROM_RAW.sh |
| Regression tests | 23/23 passed in each final run | CoMERC_LS073_v0.73_Execution/ evidence/final_release_validation/ PYTEST_FINAL_A.log; PYTEST_FINAL_B.log |
| Package file integrity | Manifest, size, SHA-256, required-file, and count checks passed | VERIFY_PACKAGE.py; MANIFESTS/SHA256_MANIFEST.json |
| Full-computation entry point | Available; writes to a new output tree and does not overwrite formal records | CoMERC_LS073_v0.73_Execution/ FULL_RECOMPUTE_ALL_CASES.sh |
| Secondary structure-dose summaries | Recomputed from all archived raw evaluation and stress records | PUBLICATION_SUPPORT/ SECONDARY_STRUCTURE_DOSE_METRICS.json; VALIDATION_V26/ recompute_secondary_structure_dose_metrics.py |

One archived historical audit script retains an old absolute path and is preserved only as provenance. The current reproduction entry points listed above use package-relative paths. A complete versioned reproducibility archive is available from the author upon request. Any later public repository record will preserve the frozen scientific files and document its license separately.

**S13. Prespecified secondary synthetic structure-dose summaries**

The statistical analysis plan prespecified mass-weighted synthetic target D95-like, target mean-dose, organ-at-risk mean-dose, and organ-at-risk D2-like errors as secondary descriptive summaries.

Table S9 reports absolute errors reconstructed from the archived case records for fixed levels, released CoMERC outputs, and released shifted-generator outputs. No value in this table was used for calibration, release decisions, or a primary gate.

**Table S9. Prespecified secondary synthetic structure-dose absolute errors (Gy).**

| **Cohort or strategy** | **n** | Mass-weighted synthetic target D95-like | Mass-weighted synthetic target mean dose | Mass-weighted synthetic OAR mean dose | Mass-weighted synthetic OAR D2-like |
|---|---|---|---|---|---|
| Primary, always NQMC-8 | 400 | 0.0456 / 0.2158 / 0.4946 | 0.0172 / 0.0577 / 0.1308 | 0.0123 / 0.0626 / 0.3123 | 0.0771 / 0.2889 / 0.9586 |
| Primary, released CoMERC outputs | 388 | 0.0208 / 0.1497 / 0.4847 | 0.0069 / 0.0422 / 0.1168 | 0.0062 / 0.0358 / 0.1447 | 0.0314 / 0.2051 / 0.5105 |
| Primary, always NQMC-16 | 400 | 0.0103 / 0.0439 / 0.1013 | 0.0039 / 0.0135 / 0.0282 | 0.0024 / 0.0142 / 0.0325 | 0.0138 / 0.0605 / 0.1896 |
| Shifted generator, released CoMERC outputs | 113 | 0.0188 / 0.1067 / 0.2042 | 0.0048 / 0.0355 / 0.0560 | 0.0041 / 0.0297 / 0.0468 | 0.0309 / 0.1389 / 0.4396 |

Cells report median / 95th percentile / maximum absolute error. CoMERC rows include released outputs only. These prespecified mass-weighted digital-structure metrics were reconstructed from archived raw records and are descriptive; they are not standard volume-based clinical DVH endpoints or clinical action thresholds. Abbreviations: CoMERC, conformal multiresolution error control; D2-like, mass-weighted dose threshold exceeded by the hottest 2% of structure mass; D95-like, mass-weighted dose threshold received by 95% of structure mass; NQMC, nested randomized quasi-Monte Carlo; OAR, organ at risk.

**S14. Post-lock controller-mechanism diagnostics**

The following diagnostics were computed after the confirmatory verdict was fixed. They were not used to tune the score, floors, tolerances, release rule, or primary analysis. They characterize how the frozen controller behaved in the archived cases and do not establish that any component is unnecessary outside the evaluated generators.

**Table S10. Post-lock controller-mechanism summary.**

| Diagnostic | Observed result | Interpretation |
|---|---|---|
| Assigned primary score driver | Eligible-cell maximum 184/400; high-gradient RMSE 111/400; density-gradient-threshold RMSE 71/400; global RMSE 34/400 | The formally joint controller was empirically maximum-error dominated. |
| Level-8 escalation | 210 escalations; maximum endpoint participated in 190 and was the only trigger in 91; all 22 actual level-8 tolerance violations were maximum-only and escalated | Escalation intercepted every observed level-8 violation but was conservative. |
| Terminal ABSTAIN | 12; all triggered only by the level-16 eligible-cell maximum bound; all 12 archived level-16 fields were within tolerance | ABSTAIN preserved terminal qualification but prevented no observed level-16 violation in this cohort. |
| Discrepancy winner | Inter-resolution 3,161/3,200; directional same-level replicate 39/3,200 | Inter-resolution discrepancy supplied nearly all effective bounds. |
| Denominator floors | 0 activations among 6,080 formal level-endpoint records | The floors did not alter any archived formal decision. |
| Density-gradient mask | Median 95.89% of eligible primary cells; >90% in 366/400; >=95% in 243/400; 100% in 11/400; correlation with global RMSE r = 0.97065 (L8) and 0.97267 (L16) | This endpoint was broad and highly redundant with global RMSE in the tested generator. |

Counts are descriptive and use the archived frozen records. L8 and L16 denote QMC sampling levels 8 and 16.

**Table S11. Primary joint-conformal coverage-miss events.**

| Case | Level | Endpoint | Error / bound | Final action | Interpretive note |
|---|---|---|---|---|---|
| CM73-E0049 | 16 | DG-RMSE | 1.049 | Release L8 | Miss occurred at the unselected level; selected output remained within tolerance. |
| CM73-E0124 | 16 | Eligible-cell maximum | 1.062 | Release L16 | Selected output remained within tolerance. |

| **Case** | **Level** | **Endpoint** | **Error / bound** | **Final action** | Interpretive note |
|---|---|---|---|---|---|
| CM73-E0131 | 16 | Eligible-cell maximum | 1.031 | Release L16 | Selected output remained within tolerance. |
| CM73-E0162 | 8 | Eligible-cell maximum | 1.019 | ABSTAIN | Archived terminal field remained within tolerance. |
| CM73-E0189 | 16 | High-gradient RMSE | 1.113 | Release L16 | Selected output remained within tolerance. |
| CM73-E0189 | 16 | Eligible-cell maximum | 2.126 | Release L16 | Largest primary error-to-bound ratio; selected output remained within tolerance. |
| CM73-E0204 | 8 | Eligible-cell maximum | 1.043 | ABSTAIN | Archived terminal field remained within tolerance. |

DG-RMSE, density-gradient-threshold RMSE (locked record field: interface RMSE). A coverage miss means that at least one prespecified error exceeded its reported conformal bound; it does not necessarily mean that the selected output exceeded its numerical tolerance.

**Table S12. Detailed shifted-generator tolerance-violating release.**

| **Parameter** | **Value** |
|---|---|
| Case / deformation | CM73-S0080 / affine + sine |
| Selected level / endpoint | 16 / eligible-cell maximum absolute error |
| Realized error / tolerance | 1.646299 Gy / 1.500000 Gy |
| Tolerance exceedance | 9.75% |
| Reported bound / error-to-bound ratio | 0.393584 Gy / 4.18 |
| Inter-resolution / replicate discrepancy | 0.554841 Gy / 0.529517 Gy |
| Eligible / high-gradient / density-gradient-mask cells | 290 / 133 / 285 |
| Numerical-reference quality control | Passed; no precision escalation |

The shifted generator was outside the target population for the calibration guarantee. This case-level description is post-lock and descriptive.

**Supplementary Figure Legends**

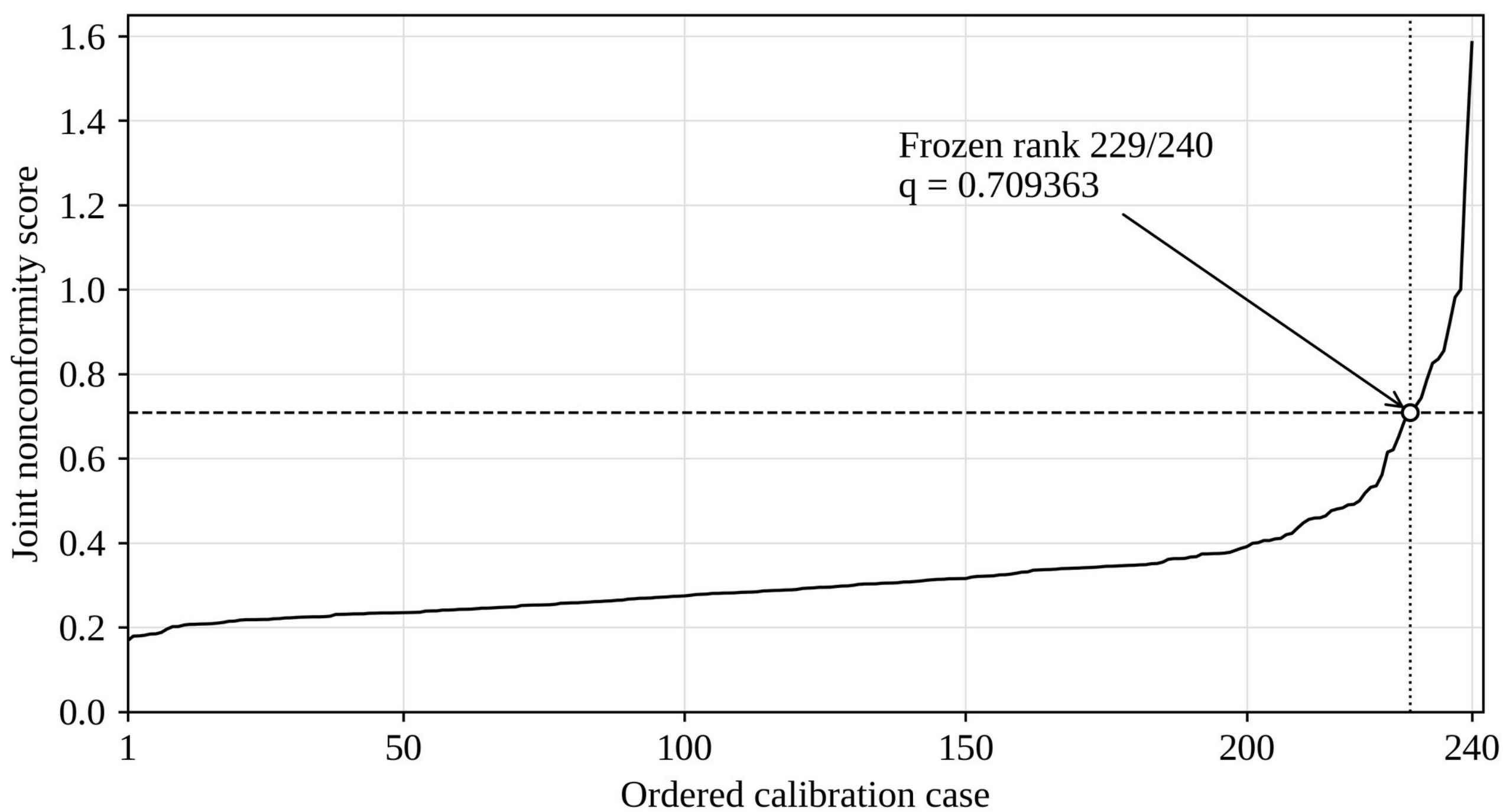

Figure S1. Calibration-score distribution. The 240 case scores are shown with the locked noninterpolated 229th order statistic, q = 0.7093632618441885.

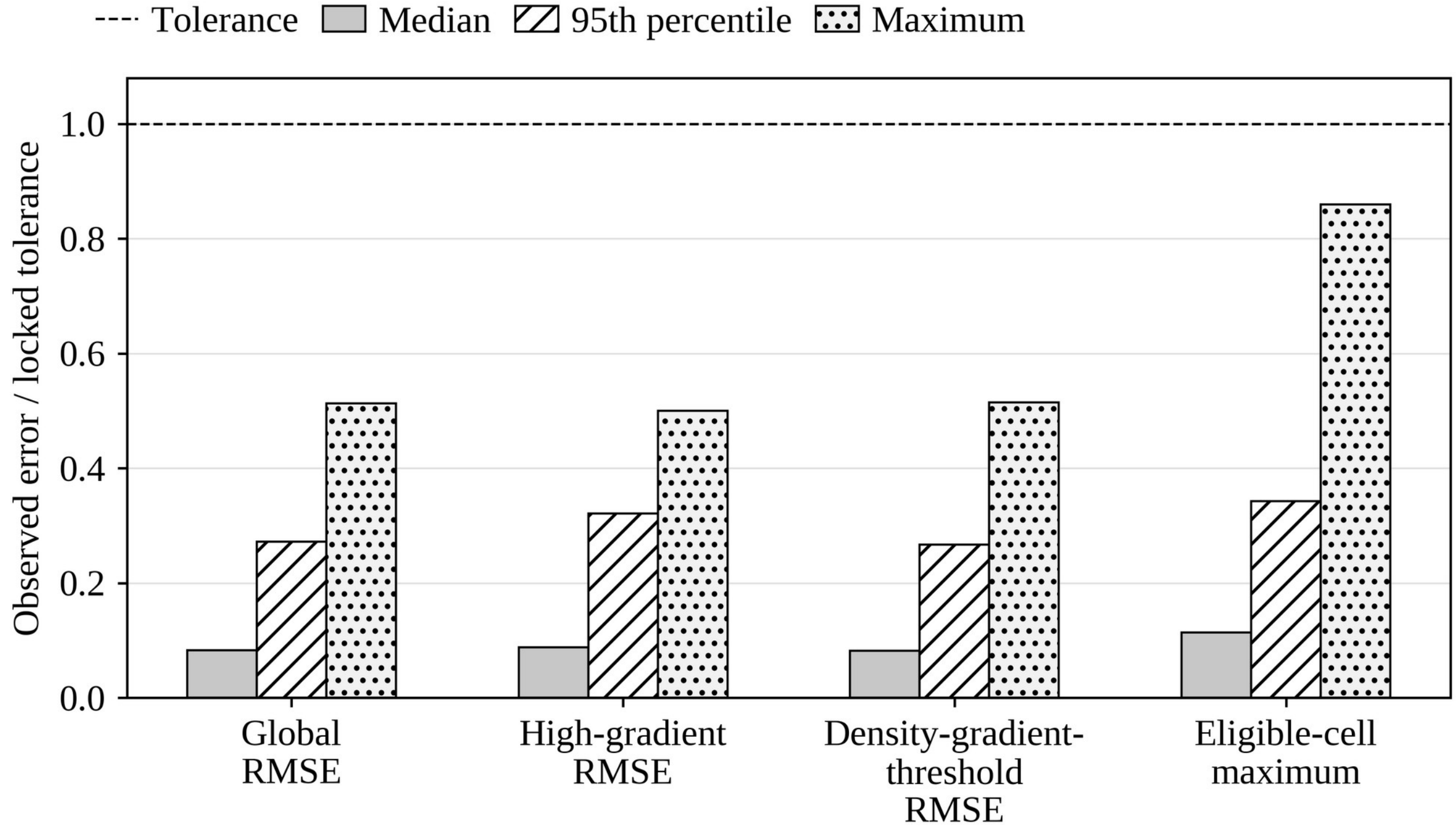

Figure S2. Released primary-output errors. Distributions of the four selected-output endpoints among the 388 released primary cases are shown relative to the corresponding numerical tolerances. The third endpoint is the density-gradient-threshold RMSE (locked record field: interface RMSE). No released primary output exceeded any tolerance.

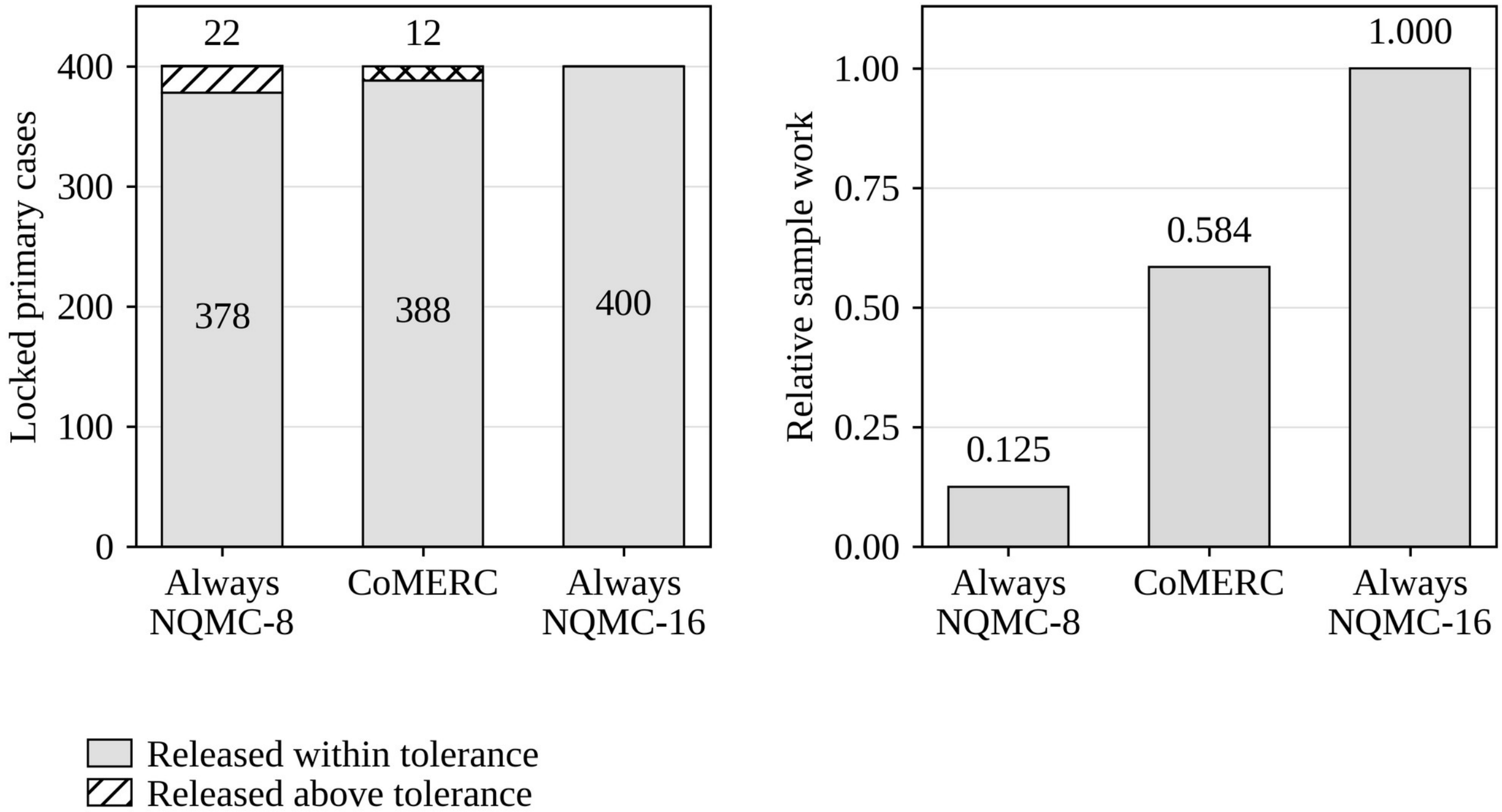

Figure S3. Same-case post-lock strategy trade-off. (A) Output outcomes for always NQMC-8, the frozen CoMERC policy, and always NQMC-16 on the same 400 primary cases. (B) Mean algorithmic sample work normalized to always executing both randomized production replicates through NQMC-16. The comparison is descriptive and was not used to tune the locked policy. CoMERC, conformal multiresolution error control; NQMC, nested randomized quasi-Monte Carlo.